\documentclass[12pt,twoside]{article}
\def\greaterthansquiggle{\raise.3ex\hbox{$>$\kern-.75em\lower1ex\hbox{$\sim$}}}
\def\lessthansquiggle{\raise.3ex\hbox{$<$\kern-.75em\lower1ex\hbox{$\sim$}}}

\newcommand{\beq}{\begin{equation}}
\newcommand{\eeq}{\end{equation}}
\newcommand{\beqa}{\begin{eqnarray}}
\newcommand{\eeqa}{\end{eqnarray}}
\newcommand{\beqan}{\begin{eqnarray*}}
\newcommand{\eeqan}{\end{eqnarray*}}
\newcommand{\ba}{\begin{array}}
\newcommand{\ea}{\end{array}}
\newcommand{\no}{\nonumber}

\newcommand{\sgn}{{\rm sgn}}
\newcommand{\Det}{{\rm Det}\,}
\newcommand{\sh}{{\rm sh}}

\newcommand{\ol}{\overline}
\newcommand{\ra}{\rightarrow}

\newcommand{\ve}{\varepsilon}

\newcommand{\wt}{\widetilde}

\newcommand{\A}{{\cal A}}
\newcommand{\C}{{\cal C}}
\newcommand{\Ha}{{\cal H}}

\newcommand{\dsum}{\displaystyle \sum}
\newcommand{\dprod}{\displaystyle \prod}

\def\nz{\ifmmode {I\hskip -3pt N} \else {\hbox {$I\hskip -3pt N$}}\fi}
\def\zz{\ifmmode {Z\hskip -4.8pt Z} \else
       {\hbox {$Z\hskip -4.8pt Z$}}\fi}
\def\qz{\ifmmode {Q\hskip -5.0pt\vrule height6.0pt depth 0pt
       \hskip 6pt} \else {\hbox
       {$Q\hskip -5.0pt\vrule height6.0pt depth 0pt\hskip 6pt$}}\fi}
\def\rz{\ifmmode {I\hskip -3pt R} \else {\hbox {$I\hskip -3pt R$}}\fi}
\def\cz{\ifmmode {C\hskip -4.8pt\vrule height5.8pt\hskip 6.3pt} \else
       {\hbox {$C\hskip -4.8pt\vrule height5.8pt\hskip 6.3pt$}}\fi}
\def\au{{\setbox0=\hbox{\lower1.36775ex%
\hbox{''}\kern-.05em}\dp0=.36775ex\hskip0pt\box0}}
\def\ao{{}\kern-.10em\hbox{``}}
\def\lint{\int\limits}
\begin{document}
\begin{titlepage}
\begin{flushright}
UWThPh-2000-39\\
hep-th/0010030\\
ESI-944-2000\\
October 5, 2000
\end{flushright}

\vspace*{1.8cm}
\begin{center}
{\Large \bf Laughlin type wave function \\[7pt]
for two-dimensional anyon fields \\[10pt]
in a KMS-state $^\star$ }\\[36pt]

 N. Ilieva$^{\ast,\sharp}$ and W. Thirring\\ [16pt]

Institut f\"ur Theoretische Physik \\ Universit\"at Wien\\
%Boltzmanngasse 5, A-1090 Wien \\
\smallskip
and \\
\smallskip
Erwin Schr\"odinger International Institute\\ for Mathematical
Physics\\

%\vfill
\vspace{0.4cm}

\begin{abstract}
The correlation functions of two-dimensional anyon fields in a
KMS-state are studied. For $T=0$ the $n$-particle wave functions 
of noncanonical fermions of level $\alpha$, $\alpha$ odd, are shown 
to be of Laughlin type of order $\alpha$. For $T>0$ they are given by 
a simple finite-temperature generalization of Laughlin's wave function. 
This relates the first and second quantized pictures of the fractional 
quantum Hall effect.

\vspace{0.8cm}
PACS numbers: 03.70.+k, 11.10.Kk, 11.10.Wx, 71.10.Pm

\smallskip
Keywords: fractional statistics, noncanonical fermions, thermal
correlators, \\
\hspace{1.9cm}quantum Hall effect

\end{abstract}
\end{center}

\vfill
\vspace{0.4cm}

{\footnotesize

$^\star$ Work supported in part by ``Fonds zur F\"orderung der
wissenschaftlichen Forschung in \"Osterreich" under grant
P11287--PHY;

$^\ast$ On leave from Institute for Nuclear Research and Nuclear
Energy, Bulgarian Academy of Sciences, Boul.Tzarigradsko Chaussee
72, 1784 Sofia, Bulgaria

$^\sharp$ E--mail address: ilieva@ap.univie.ac.at}
\end{titlepage}
\vfill \eject

\setcounter{page}{2}
The ingenious suggestion by Laughlin \cite{La} of a wave function 
describing the fractional quantum Hall effect \cite{Nob} is based on an
empiric analysis of the approximate effective theory of this phenomenon. 
It is to be expected that a microscopic derivation of this function shoud 
rest upon a (second-quantized) picture of the edge excitations that form 
the corresponding chiral Luttinger liquid \cite{W} --- the quasiparticles 
and quasiholes characterized by fractional statistics \cite{Hal} (see also 
\cite{FW} and references therein). The anyon fields constructed in 
\cite{epj, tmp, INT} share most of the features of the above objects. 
Their exchange relations, the $\alpha$-commutators, show, however, a 
temperature dependence, so that finite-temperature effects in this 
construction manifest themselves not only globally but also locally.
The important point is that their correlation functions do not factorize in 
the conventional manner. As we shall see, the wave function of an
$n$-particle $\alpha$-anyon state is given by the $\alpha$-power of a
Slater determinant and for zero temperature is of Laughlin type of order 
$\alpha$. 

\vspace{12pt}
In the discussion of Bose--Fermi duality at finite temperature there 
appear in a natural way field operators with exotic exchange relations, 
namely anyons \cite{epj, tmp}
\beq
\Psi_\alpha(x) := \lim_{\ve \ra 0} n_\alpha(\ve) \exp \left[ i
2\pi\sqrt{\alpha} \int_{-\infty}^\infty dy \; \phi_\ve(x-y) j(y)
\right]\, ,\quad \alpha\in{\bf R}^+\,,
\eeq
with $n_\alpha$ --- some renormalization parameter
and $\phi_\ve (x)$ --- an approximation to the Heaviside function
$$
\lim_{\ve\ra 0}\phi_\ve(x) = \Theta(x), \quad \phi_\ve(x) \in
H_1\, ,
$$
where $H_1$ is the Sobolev space, $H_1 = \{f: f, f' \in L^2 \}$. It 
should be emphasized that these are one-dimensional objects in
contrast to the ``conventional" anyons, which exist in two spatial 
dimensions (compare \cite{FW}). Their anyonic
nature is expressed through the exotic exchange relations they
satisfy \cite{tmp, INT}, which interpolate between bosons and fermions. An 
algebraic approach to fractional statistics in one dimension was 
introduced in \cite{LM}, based on the analysis of the 
Heisenberg--Weyl algebra of observables for identical particles. 

In our context, the anyon fields (1) appear as follows: 
If we start with bare fermion fields $\{\psi(x), \psi^*(y)\} = \delta (x-y)$,
the smeared fields $\psi_f:=\int dx f(x)\psi(x)$ are bounded operators,
$\Vert \psi_f\Vert = \int dx \vert f(x)\vert^2$, and form the 
canonical-anticommutation-relations (CAR) algebra $\A\,$. The norm of
$\psi_f$ tends to infinity if $f(x)\ra\delta(x-x_0)$ and thus the current 
$j(x_0) = \psi^*(x_0)\psi(x_0)$ cannot be reached as a norm limit. However, 
in a particular representation $\pi$ the limit $f(x)\ra\delta(x-x_0)$ for the 
normal-ordered product 
$:\!\psi^*_f\psi_f\!:$ can exist as a strong limit. This was shown to happen 
in  \cite{epj} in the representation $\pi_\beta$ given by the equilibrium 
(KMS) state for a temperature $T=1/\beta$. This means that if one enlarges 
$\pi_\beta(\A)$ by adding all strong limits, the ensuing algebra (denoted 
by $\pi_\beta(\A)''$) will already contain the bosonic algebra $\A_c$ 
spanned by the currents $j(x)$. Its structure is the same for all non-negative 
temperatures $0\leq T<\infty$ but changes for $\beta=0$ or $\beta < 0$. With 
these currents one can try to construct anyon fields by (1) but this works 
only by adding in addition to strong limits some ideal elements to form a 
bigger algebra $\bar \A_c\,$. The representation $\pi_\beta$ extends naturally 
to it to give $\bar \pi_\beta(\bar\A_c)$ where one might again include strong 
limits to get $\bar \pi_\beta(\bar\A_c)''$. In this v. Neumann algebra  Fermi 
fields like the ones  proposed by Mandelstam \cite{Man} can be identified,
thus giving rise to another Fermi  algebra, CAR (dressed). Thus  the 
phenomenon of the Bose--Fermi duality can be interpreted as the existence 
of this particular chain of algebraic inclusions \cite{epj}
$$
{\rm CAR}({\it bare}) \subset \pi_\beta(\A)'' \supset\A_c \subset \bar
\A_c \subset \bar\pi_\beta(\bar\A_c)'' \supset {\rm CAR}({\it dressed})\,.
$$

The crucial ingredient needed at the first step is the appropriately chosen 
state. We choose the KMS-state which is unique for the shift (which for 
chiral fields is equivalent to the time development) over the CAR algebra. 
A limiting case would be to chose the Dirac vacuum by filling all 
negative energy levels in the Dirac sea. This is  what has originally been 
done in the thirties \cite{J,BNN}, in the attempts for constructing a neutrino 
theory of light and recovered later by Mattis and Lieb \cite{ML} in the 
context of the Luttinger model.

The essential result is the appearance of an anomalous (Schwinger)
term in the quantum current commutator
\beq
[j(x),j(x')] = -\,\frac{i}{2\pi}\,\delta'(x\!-\!x')\,.
\eeq
For the smeared currents one gets
\beq
[j_f, j_g] = \int_{-\infty}^\infty \frac{dp}{(2\pi)^2} \; p \wt f(p) \wt
g(-p) =\frac{i}{2\pi} \int_{-\infty}^\infty dx f'(x)g(x)
 = i\sigma(f,g)\,,
\eeq
$\sigma(f,g)$ being the symplectic form on the current algebra $\A_c\,$.
However, an important detail might be overseen that way:
symplectic structure (3) though formally independent on $\beta$
(see also \cite{HG}),  for $\beta < 0$ changes its sign, $\sigma
\to -\sigma$, and for $\beta = 0$ (the tracial state) becomes
zero.
Note that it is the parity $P$ (which suffers a destruction on the passage 
from the CAR-algebra to the current algebra \cite{dub, tmp}) that
relates the states corresponding to positive and negative temperatures
$$
\omega_{-\beta} = \omega_\beta \circ P\,.
$$

\medskip

However, the dressed fermions present in $\bar \pi_\beta(\bar\A_c)''$ 
are only a special type of anyons, defined by a particular value of the 
statistic parameter $\alpha$, namely $\alpha = 1$.  In \cite{INT} the 
general anyonic field
\beq
\Psi_\alpha (x) \simeq e^{\,i 2\pi\sqrt{\alpha}\lint_{-\infty}^{x}j(y)dy}
\eeq
has been represented as an operator valued distribution in a Hilbert space
by exhibiting its
$n$-point function in a $\tau$-KMS state $\omega$. Thus completed, the
rigorous construction of the anyonic fields proposed in \cite{epj, tmp}
allows for a detailed analysis of various properties of these
interesting objects, in particular their thermal behaviour and its
relevance for the corresponding correlation functions.

The field $\Psi_\alpha(x)$ (4) is both infrared and ultraviolet singular.
The infrared divergence amounts to the fact that admitting the (smeared)
step function as a test function, one creates new elements in the field
algebra which lead to orthogonal sectors in a larger Hilbert space
\beq
\bar\Ha_\beta = \oplus \bar\Ha^n_\beta, \qquad
\bar\Ha^n_\beta = \A_c\,\dprod_{i=1}^{n}
\Psi_{\alpha}(x_i)\vert\Omega\rangle\,.
\eeq
The ultraviolet divergence is of another type: it does not lead out
of $\bar\pi_\beta$ if we smear $j(y)$ over a region of size $\eta$ to get
$\Psi_{\alpha,\eta}$ and consider the renormalized field
\beq
\lim_{\eta\ra 0^+}c_\alpha(\eta)\int dx f(x)\Psi_{\alpha,\eta}(x) =
\Psi_\alpha(f) \,
\eeq
with a suitable $c_\alpha(\eta)$. This limit exists in a strong sense and
$\Psi(f)$ has finite $n$-point functions \cite{INT}.

As already mentioned, for particular values of the statistic parameter 
$\alpha$ some special families of such renormalized field operators are 
distinguished: for odd $\alpha$'s we get fermions and for even 
$\alpha$'s  --- bosons. However, only the field $\Psi_1$  turns out  to 
be a canonical Fermi field,
$$
[\Psi_1^*(x), \Psi_1(x')]_+ = \delta (x-x')\,,
$$
with an $n$-point function of the familiar determinant form.
$\Psi_2$  is a non-canonical Bose field,
whose commutator is not a $c$-number$$
[\Psi_2^*(x), \Psi_2(x')] \simeq \delta' (x-x') + ij(x)\delta(x-x')\, .
$$
Similarly, the operator $\Psi_3$ describes a non-canonical  (unbounded)
Fermi field. For $\alpha\not\in {\bf Z}$ the anyonic commutator vanishes.

Investigation of anyonic field operators of the type (1),(4) represents
by far not only an academic interest --- such fields might become of
importance in solid-state physics, in problems like quantum wires and
fractional quantum Hall effect (FQHE). The relation between the objects
there involved and the field operators (1) is rather obvious \cite{I}.

Thus, in quantum Hall fluids, the edge-excitation operators are identified 
with Wen's fermions \cite{W, FK} and have an exotic statistics  depending 
on the filling fraction $\nu=\alpha^{-1}$. However, in the fermionic
case --- $\alpha = 2n+1$, so for Laughlin's states --- one has to distinguish
between fermions, corresponding to $n=0$ and $n\not=0$. As just
mentioned, these fields, though locally anticommuting, are quite different:
the former are canonical fields, while the latter are not and this difference
shows up also in their thermal properties.

\vspace{12pt}
The current algebra $\A_c$ is defined for instance for $j_f$'s with $\, f
\in C_0^\infty$, that is with functions which vanish for
$x\ra\pm\infty$. The anyons (4) are also Weyl operators but for which
the smearing function is $f_x^\alpha(y) = 2\pi\sqrt{\alpha}\,\Theta(x-y)$.
The structure of $\A_c$ is determined by the symplectic form $\sigma(f,g)$
(3) which is actually well defined  for the Sobolev space,
$\sigma(f,g) \ra \sigma(\bar f, \bar g), \, \bar f, \bar g \in H_1, \, H_1
= \{f : f,f' \in L^2\}\,$. The state $\bar \omega_\beta$ can be
extended to $H_1$ as well, since $\,\bar \omega_\beta(e^{ij_{\bar f}}) >
0\,$ for $\,\bar f \in H_1$.  Thus, the symplectic form (3) may be given
a sense for functions that tend to a constant, however they cannot be
reached as limits of functions from $\C_0^\infty$. Let $\Phi_{x,\delta}$ 
be such a function, with $\delta$ being the infrared-regularization 
parameter. The point is then that
$\sigma(\Phi_{x,\delta}, \Phi_{x',\delta'})$  depends
on the order in which the limits $\delta, \delta' \ra\infty$ are taken and
only for $\delta= \delta'\ra\infty$ we get the desired result
$i\,\sgn(x-x')$. Since this appears in the $c$-number part, in no
representation can $j(\Phi_{x,\delta})$ converge strongly. Nevertheless,
for functions with the same (nontrivial) asymptotics at, say, $x\ra\infty$
and whose difference $\in h$ (see below) one can succeed in getting the
expectation values as limits.

Recall that for Weyl operators relation (2) is replaced by the
multiplication law
\beq
e^{ij(f)} \; e^{ij(g)} =
e^{\frac{i}{2} \sigma(g,f)} \; e^{ij(f+g)}\,.
\eeq
The $\tau$-KMS states are translation-invariant equilibrium states
at an inverse temperature $\beta$. On $\A_c$ such a state is given by
the two-point function
$$
\omega(j(f)j(g)) = \int dxdy \,K(x-y)f(x)g(y)\,
$$
with a kernel
\beq
K(x-y) =-\lim_{\ve\ra0^+}\,\frac{1}{(2\pi)^2\,\sh^2(x-y-i\ve)}\, .
\eeq
The expectation of the Weyl operators is given by
\beq
\omega(e^{ij(f)}) = e^{-{1\over 2}\langle f | f\rangle}\, ,
\eeq
where the scalar product $\langle f | g\rangle$ defines the one-particle
real Hilbert space $h$ of the $f$'s. For consistency, it has to satisfy
$$
i\sigma(f,g) = \left(\langle g|f\rangle - \langle f|g\rangle\right)\, .
$$
Eqs.(7),(9) imply
$$
\omega(e^{ij(f)}e^{ij(g)}) = \exp{\left\{-{1\over 2}
\left[ \langle f | f \rangle+\langle g | g \rangle +
2\langle f | g \rangle \right]\right\}}\, ,
$$
or generally
\beq
\omega(\prod_k e^{ij(f_k)}) = \exp{\left\{-{1\over 2}
\left[\dsum_k \langle f_k | f_k \rangle +
2\dsum_{k<m} \langle f_k | f_m \rangle \right]\right\}}\, .
\eeq

A trivial integration then yields for the $\alpha$ two-point function
\cite{tmp, INT}
\beq
\omega(\Psi^*_\alpha(x)\Psi_\alpha(y)) = \omega(e^{-ij(f_{x,\ve}^\alpha)}
e^{ij(f_{y,\ve}^\alpha)}) 
= \left(\frac{i}{2\beta\sh{\frac{\pi(x-x'-i\ve)}{\beta}}}\right)^{\alpha}\,.
\eeq

For all $\alpha$'s the two-point function (for $x > x'$ and $\beta =
\pi$)
\beq
\langle \Psi_\alpha^*(x) \Psi_\alpha(x') \rangle_\beta =
\langle \Psi_\alpha(x) \Psi_\alpha^*(x') \rangle_\beta =
\left(\frac{i}{2\pi \sh(x-x')}\right)^{\alpha} =: S_\alpha(x-x')
\eeq
has the desired properties
\begin{description}
\item [(i)] {\it Hermiticity:}
$$
S_\alpha^*(x) = S_\alpha(-x) \, \Longleftrightarrow \, \langle
\Psi_\alpha^*(x)\Psi_\alpha(x')\rangle_\beta^* = \langle
\Psi_\alpha^*(x')\Psi_\alpha(x)\rangle_\beta\,;
$$
\item [(ii)] {\it $\alpha$-commutativity:}
$$
S_\alpha(-x) = e^{i\pi\alpha}S_\alpha(x) \, \Longleftrightarrow \,
\langle\Psi_\alpha(x')\Psi_\alpha^*(x)\rangle_\beta =
e^{i\pi\alpha}\langle\Psi_\alpha^*(x)\Psi_\alpha(x')\rangle_\beta \, ;
$$
\item [(iii)] {\it KMS-property:}
$$
S_\alpha(x) = S_\alpha(-x+i\pi) \, \Longleftrightarrow \,
\langle\Psi_\alpha^*(x)\Psi_\alpha(x')\rangle_\beta =
\langle\Psi_\alpha(x')\Psi_\alpha^*(x+i\pi)\rangle_\beta\, .
$$
\end{description}

For $\alpha = 2$ and an arbitrary temperature $\beta^{-1}$
we get like for the $j$'s
\beq
\langle\Psi_{2}^*(x)\Psi_{2}(x')\rangle_\beta =
-\frac{1}{\left(2\beta\,\sh{\frac{\pi(x-x'-i\ve)}{\beta}}\right)^2}\, ,
\eeq
similarly, for $\alpha = 3$ we get a different kind of fermions
\beq
\langle\Psi_{3}^*(x)\Psi_{3}(x')\rangle_\beta =
-\frac{i}{\left(2\beta\,\sh{\frac{\pi(x-x'-i\ve)}{\beta}}\right)^3}\, .
\eeq
These fields, though locally (anti)commuting,  are not canonical and
this becomes transparent by analysing temperature dependence and
operator structure of their exchange relations. However, the Fermi fields
$\Psi_{2n+1}$ are similar to Wen's fermions
$$
\langle \psi(z)\psi^\dag(w)\rangle \sim
\frac{1}{(z-w)^{2n+1}}
$$
that correspond to Laughlin's plateaux in the theory of the FQHE
(considered at a finite temperature), in which case these construction
would provide a second-quantization picture of this phenomenon. For a
detailed analysis of this relation we refer to \cite{IT}.

For the $n$-point function to get something finite for $\delta\ra\infty$
we have to take operators of the form
$$
\prod e^{\pm ij(f_{x_k}^\alpha)} := \prod e^{ij(\ol f_{x_k}^\alpha)}\, ,
$$
where $\ol f_{x_k}^\alpha = s_k^\alpha f_{x_k}\,$, $\,s_k^\alpha = \pm
\sqrt\alpha$. Since the individual expressions diverge with
$\delta\ra\infty$, if the anyon contributions do not neutralize, i.e. if
$\sum_k s_k^\alpha \not=0$, the exponent as a whole diverges, so the 
expectation value (10) vanishes.  If, on the other hand, 
$ \sum_k s_k^\alpha=0$, those terms that contain $\delta$
can be combined in pairs to cancel and one thus remains
in the limit $\delta\ra\infty$ (with the normalization factors $c_\alpha$ 
(6) taken into account) with
\beqa
&&\omega\left(\Psi_\alpha^*(x_1)\dots\Psi_\alpha^*(x_n)
\Psi_\alpha(y_n)\dots\Psi_\alpha(y_1)\right) \no\\[8pt]
& = &
\frac{\dprod_{k>l}(\sh(x_k-x_l-i\ve))^\alpha\dprod_{k>l}
(\sh(y_k-y_l-i\ve))^\alpha}
{\dprod_{k,l} \left(-2\pi i \,\sh(x_k-y_l-i\ve)\right)^{\alpha}}\,.
\eeqa

For the case $\alpha = 1$  with the help of Cauchy's determinant
formula the $n$-point function (15) can be rewritten as \cite{INT}
\beq
 \frac{\dprod_{i>k}
\sh(x_i-x_k-i\ve)\dprod_{i>k}\sh(y_i-y_k-i\ve)}
{\dprod_{i,k}\sh(x_i-y_k-i\ve)}= \Det\frac{1}{\sh(x_i-y_k-i\ve)} \,.
\eeq
We do not prove (16) but just remark that  the two expressions have 
the same pole structure, homogeneity degree and symmetries. The 
state over the field algebra $\A_1$ is quasifree, the fermion two-point 
function being given by
$$
\omega(\Psi_1^*(x)\Psi_1(y)) = \frac{i}{2\pi\,\sh(x-y-i\ve)}\, .
$$
It satisfies the KMS condition with respect to the shift for temperature
$\beta=\pi$. For arbitrary temperature by scaling arguments it follows
\beq
\omega_\beta(\Psi^*_1(x)\Psi_{1}(y)) =
\frac{i}{2\beta\,\sh \frac{\pi(x-x'-i\ve)}{\beta}}\, .
\eeq
Since the $\tau$-KMS state over the CAR-algebra is unique, we thus recover
the original free fermions.

Evidently, for $\alpha\not=1$ the state is determined again by the
two-point function but not in a way that corresponds to a truncation.
Recall that an $n$-particle state is given by
$$
|n\rangle = \int \Psi^*(x_1)\dots\Psi^*(x_n)|\Omega\rangle F(x_1,\dots,
x_n)dx_1\dots dx_n\,,
$$
its wave function being
$$
\phi(x_1,\dots, x_n) := \langle\Omega|\Psi(x_1)\dots\Psi(x_n)|n\rangle \,.
$$
$|n\rangle$ is a Slater state if $F(x_1,\dots, x_n) = \prod_{i}f_i(x_i)$, for
which $\phi$ is a determinant if $|\Omega\rangle$ is the Fock vacuum.
We shall call a wave function $\phi$ being of {\it Laughlin type of
order $\alpha$}, if $\phi$ is of the form $\prod_{i>k}(x_i-x_k)^\alpha\prod_m
\Phi(x_m)$, for $0<|\Phi|<\infty$ and $\alpha$ odd.

\smallskip
Note that because of the anti-commutativity of the $\Psi$'s, the Slater
determinant $F = \Det f_i(x_j)$ gives for fermions the same state as 
$\prod_{i}f_i(x_i)$. If $|\Omega\rangle$ is the vacuum then $|n\rangle = 0$ 
if for some $f_k$, $\mbox{supp } \tilde f_k \subset (0, -\infty)$. 
Furthermore, for functions $f$ with $\mbox{supp } \tilde f_k 
\subset (0, \infty)$ such that $f(x)$ is analytic in the upper half-plane,
the set $\left\{f_z(x)=(x-z)^{-1}, \mbox{Im }z < 0 \right\}$ is total, i.e. 
their linear combinations are dense. Therefore we get for $\beta\ra\infty$ 
up to a normalization factor
\beqan
\phi(x_1,\dots, x_n) & = &\prod_{i>j}(x_i-x_j)^\alpha
\int\frac{dy_1}{(y_1-z_1)}\dots
\frac{dy_n}{(y_n-z_n)}\,\frac{\dprod_{k>l}(y_k-y_l)^\alpha }
{\dprod_{k,l}(x_k-y_l+i\ve)^\alpha}\\[6pt]
& = & \frac{\dprod_{l>j}(x_l-x_j)^\alpha \dprod_{k>l}(z_k-z_l)^\alpha}
{\dprod_{k,l}(x_k-z_l+i\ve)^\alpha}\, ,
\eeqan
which is exactly the desired Laughlin-type wave function with
$$
\Phi(x) = \prod_l (x-z_l+i\ve)^{-\alpha}
$$
with all required properties. Thus, 
{\it for fermions of order $\alpha$ a Slater state in vacuum has
Laughlin-type wave function of order $\alpha$ for a total set of $f$'s.}

Obviously, for $\alpha = 1$ $\phi$ is (up to a constant factor) the Slater
determinant $\Det\left(x_k^{\,j} \Phi(x_k)\right)$, for other $\alpha$'s it
is the $\alpha$-power of such  a determinant.

For finite temperature $T=\beta^{-1}$, $\pi(x_l-x_k)$ is replaced by
$\beta\,\sh{[\pi (x_l-x_k)/\beta]}$ and $\Phi(x)$ --- by $\beta^{n}
\prod_{l=1}^n \sh^{-\alpha}[\pi(x-z_l+i\ve)/\beta]$. By pulling out
$\prod_{l>k}(x_l-x_k)^\alpha$ the rest gets a factor
$\prod_{l>k}\sh^\alpha [(x_l-x_k)/\beta]/(x_l-x_k)$ which is finite
and symmetric but no longer a pointwise product. Thus the wave 
function is not exactly of the Laughlin type.

\vspace{12pt}
To summarize, the operators $\Psi_\alpha(x)$, $\alpha\in{\bf R}^+$,
constructed in \cite{epj, tmp, INT} describe, depending on the value of
the statistic parameter $\alpha$, a variety of fields --- generally anyons 
but also (the integer classes) bosons and fermions. However, only the 
first-level fermions are canonical fields. This shows that local 
anticommutativity alone does not guarantee the uniqueness of the 
KMS-state, one needs in addition the CAR-relations. Thus 
$\Psi_\alpha$'s, $\,\alpha \in 2{\bf N}+1\,,\,$ describe an infinity of 
inequivalent fermions, characterized by temperature-dependent correlation
functions and exchange relations. This dependence means a loss of the 
local normality of the representations corresponding to different 
temperatures, hence --- (already local) observability of the temperature 
effects.

The wave functions for $T=0$ of an $n$-particle state of noncanonical 
fermions of level $\alpha$ (so, for $\alpha$ odd) are of Laughlin type of 
order $\alpha$. For the first ``extended" Fermi-class --- $\Psi_3$, the original 
Laughlin's form is obtained. For $T>0$ the wave functions are of a similar
form, which is a simple finite-temperature generalization of the 
zero-temperature case. Such a relation between our noncanonical Fermi 
fields and Wen's fermions is another argument for their detailed analysis 
because of the possibility for a second-quantization picture of the fractional
quantum Hall effect they provide. In particular, such anyon fields naturally 
appear in one of the two $(1\!+\!1)$-dimensional chiral theories whose 
tensor product gives the second quantization of the Hall Hamiltonian. 
They describe the macroscopic Hall current and the microscopic Larmor 
precession respectively \cite{IT}.

\section*{Acknowledgements}
We are grateful to A. Alekseev, L. Georgiev and  K.-H. Rehren for the
inspiring suggestions and to H. Narnhofer for the interest and critical
remarks.

N.I. acknowledges hospitality and financial  support of the International
Erwin Schr\"odinger Institute for Mathematical Physics where part of the
research has been performed. This work has been supported in part also
by ``Fonds zur F\"orderung der wissenschaftlichen Forschung in
\"Osterreich" under grant P11287--PHY.

\end{document}